\documentclass[iop,apj]{emulateapj}
\usepackage{amsmath,amssymb,amstext}
\usepackage{float}
\usepackage{lineno}
\usepackage[breaklinks,colorlinks,citecolor=blue,linkcolor=magenta]{hyperref}

\usepackage{aas_macros}
\usepackage{natbib}
\shorttitle{Evidence for Sub-Chandrasekhar Mass Type Ia Supernovae}
\slugcomment{Accepted for publication in ApJL, December 23 2017}
\shortauthors{Goldstein \& Kasen}

\newcommand{\snia}{{\rm SN~Ia}}
\newcommand{\sneia}{{\rm SNe~Ia}}
\newcommand{\q}[1]{{\tt #1}}

\newcommand{\Nifs}{\ensuremath{^{56}\mathrm{Ni}}}

\newcommand{\Cofs}{\ensuremath{^{56}\mathrm{Co}}}

\newcommand{\Feff}{\ensuremath{^{54}\mathrm{Fe}}}
\newcommand{\Crfe}{\ensuremath{^{48}\mathrm{Cr}}}
\newcommand{\Vfe}{\ensuremath{^{48}\mathrm{V}}}
\newcommand{\Tife}{\ensuremath{^{48}\mathrm{Ti}}}
\newcommand{\mej}{\ensuremath{M_{\rm Ej}}}
\newcommand{\mni}{\ensuremath{M_{^{56}\mathrm{Ni}}}}

\newcommand{\mime}{\ensuremath{M_{\rm IME}}}
\newcommand{\msun}{{\ensuremath{M_{\odot}}}}
\newcommand{\Msun}{{\ensuremath{M_{\odot}}}}

\newcommand{\Mej}{\ensuremath{M_{\mathrm{Ej}}}}

\newcommand{\Mch}{{\ensuremath{M_\mathrm{ch}}}}
\newcommand{\mch}{\Mch}

\newcommand{\dmfb}{\ensuremath{\Delta M_{15}(B)}}

\newcommand{\epth}{{\ensuremath{\epsilon_{\rm th}}}}
\newcommand{\MCcode}{SEDONA}
\newcommand{\kms}{\ensuremath{\mathrm{km~s}^{-1}}}

\newcommand{\Nmix}{\ensuremath{N_{\mathrm{mix}}}}

\newcommand{\nmod}{4,500}

\newcommand{\fni}{\ensuremath{f_{\mathrm{Ni}}}}
\newcommand{\fco}{\ensuremath{f_\mathrm{CO}}}
\usepackage{amsmath}
\begin{document}

\title{Evidence for Sub-Chandrasekhar Mass Type Ia Supernovae\\from an Extensive Survey of Radiative Transfer Models}

\author{
    Daniel~A.~Goldstein\footnotemark[1] \& 
    Daniel~Kasen
}
\affil{Department of Astronomy, University of California, Berkeley, 501 Campbell Hall, Berkeley, CA 94720, USA}
\affil{Lawrence Berkeley National Laboratory, 1 Cyclotron Road, Berkeley, CA, 94720, USA}
\footnotetext[1]{e-mail: \url{dgold@berkeley.edu}}

\begin{abstract}
  There are two classes of viable progenitors for normal Type Ia supernovae (\sneia): systems in which a white dwarf explodes at the Chandrasekhar mass (\mch), and systems in which a white dwarf explodes below the Chandrasekhar mass (sub-\mch). 
It is not clear which of these channels is dominant; observations and light curve modeling have provided evidence for both.
Here we use an extensive grid of 4,500 time-dependent, multiwavelength radiation transport simulations to show that the sub-\mch\ model can reproduce the entirety of the width-luminosity relation (WLR), while the \mch\ model can only produce the brighter events $(0.8 < \dmfb < 1.55)$, implying that fast-declining \sneia\ come from sub-\Mch\ explosions.
We do not assume a particular theoretical paradigm for the progenitor or explosion mechanism, but instead construct parameterized models that vary the mass, kinetic energy, and compositional structure of the ejecta, thereby realizing a broad range of possible outcomes of white dwarf explosions.
We provide fitting functions based on our large grid of detailed simulations that map observable properties of \sneia\ such as  peak brightness and light curve width to physical parameters such as \Nifs\ and total ejected mass.
These can be used to estimate the physical properties of observed \sneia.
\end{abstract}

\keywords{(stars:) supernovae: general}
\maketitle

\section{Introduction}
\label{sec:intro}
Type Ia supernovae (\sneia)  play a crucial role in astrophysics: they contribute to the chemical enrichment of galaxies, represent a key endpoint of stellar evolution, and continue to provide precise distances that enable the study of dark energy.
Yet we know surprisingly little about their origins.
Several lines of observational evidence, including early observations of the  nearby \snia\ 2011fe \citep{nugent11}, indicate that the events result from the runaway thermonuclear explosion of at least one accreting carbon-oxygen (C/O) stellar core in a binary system \citep{hillebrandt00, maozreview}, most likely a white dwarf (WD; \citealt{bloom12}). 
But several viable theories exist for the nature and mass of the donor star, the triggering mechanism of the explosion, and the process by which thermonuclear burning propagates through the WD. 

Potential progenitors of normal \sneia\ can  be roughly divided into two categories (see \citealt{wangandhan} or \citealt{maozreview} for recent reviews): systems that produce an explosion near the Chandrasekhar mass, the maximum mass of a stable, non-rotating WD ($\Mch\approx1.4\,\Msun$), and systems that produce an explosion below the Chandrasekhar mass (sub-\Mch). 
In the classic \mch\ scenario \citep{sd}, a C/O WD accretes mass from a non-degenerate donor, usually a main sequence star or a red giant, until it approaches \Mch\ and centrally ignites. 

A main alternative to the classic \mch\ channel is the sub-\Mch\ double-detonation scenario, in which accreted helium on the surface of a WD in a close binary detonates before the star nears \Mch, launching a shock wave that travels inward to detonate the C/O \citep{taam80,nomoto82,woosley94,shen09,fink10}.
Double-detonations can be either single-degenerate \citep[if the donor is a non-degenerate helium star;][]{livne90} or double-degenerate \citep[if the donor is another WD;][]{bildsten07,guillochon10}.

For many years, the classic \mch\ scenario was favored as the dominant \snia\ progenitor channel \citep[e.g.,][]{hillebrandt00,woosley2007}.
But recently, independent lines of evidence that disfavor this scenario as the singular path to \sneia\ have emerged.
\cite{gilfanov10} showed that the X-ray fluxes of nearby galaxies are too low for classic \Mch\ progenitors, which are X-ray bright, to account for the observed  \snia\ rate.
\cite{sim10} and \cite{woosley11} showed that sub-\mch\ models can reproduce the observed properties of \sneia\ and the basic light curve width-luminosity relation \citep[WLR;][]{phillips93}.
\cite{scalzo14a,scalzo14b} used the gray formalisms of \cite{arnett82} and \cite{jeffery99} to infer the masses of \sneia\ from their bolometric light curves, finding evidence for a significant rate of sub-\mch\ events.
Recent evidence against the classic \mch\ scenario includes the lack of observed surviving companions in nearby \snia\ remnants \citep{shappee13}, the low metallicities of nearby galaxies \citep{mcwilliam17}, the correlation between \snia\ luminosity and stellar population age \citep{shen17b}, the lack of observed emission from stripped companion material \citep{shappee16,botyanszki17b}  and simulations of spectrum synthesis for small sets of sub-\mch\ explosion models \citep{blondin17b,blondin17,shen17}, which suggest sub-\mch\ events may be required to explain  fast-declining \sneia. 

In this article,  we present detailed radiation transport simulations  of a broad grid of \nmod\ parametrized, one-dimensional (1D) supernova ejecta models.
The models are designed to capture the essential degrees of freedom in \snia\ spectrum synthesis without assuming a particular theoretical paradigm for the progenitor or explosion mechanism.
Our broad model survey suggests that fast-declining \sneia\ must have sub-\Mch\ progenitors.
This implies that the classic \mch\ scenario cannot explain all \sneia.

\section{A Simple Model for The Ejecta}
\label{sec:explosion}

The starting point for our simulations is a model describing the composition, density, and temperature of the debris produced by the thermonuclear explosion of a C/O WD. 
We assume that the explosion completely unbinds the star, and that the process of nuclear burning fuses part of the C/O into heavier elements.
The nuclear energy released by burning, the internal and gravitational binding energies of the exploding WD, and other progenitor-specific  terms that we do not model explicitly here (e.g., WD rotation; \citealt{yoon05}) control the final kinetic energy of the ejecta, which expands and reaches homology minutes after the explosion. 
The ejecta is assumed to be spherically symmetric, consistent with spectropolarimetric observations that show that global asymmetries in \sneia\ are minor \citep{polarimetry}. 

We write the kinetic energy $E_K$ of the ejecta as $E_K = \frac{1}{2}\mej v_K^2$, where \mej\ is the total ejected mass  and $v_K$ is the kinetic energy velocity, both free parameters of the model.
We let both \mej\ and $v_K$ float independently to implicitly capture progenitor-specific contributions to the kinetic energy; thus the kinetic energies of our models  span a broader range than what would be obtained by summing the contributions of nuclear burning,  internal energy, and  gravitational binding energy alone.

We adopt an exponentially decaying density profile 
\begin{equation}
\label{eq:density}
	\rho(v, t) = \rho_0 \left(\frac{t_0}{t}\right)^3 e^{-v/v_e},
\end{equation}
where $v_e$ is the characteristic $e$-folding velocity of the atmosphere,
given by
\begin{equation}
	v_e = \sqrt{\frac{E_K}{6 M_{\rm Ej}}} = \frac{v_K}{\sqrt{12}}.
\end{equation}
This profile has provided a good fit to the results of many explosion simulations \citep{woosley2007}, but it has been known to overpredict the central densities of the ejecta of sub-\mch\ explosions \citep{shen17}.
In Section \ref{sec:discussion}, we study the effects of  a less centrally concentrated density profile. 

In a detailed calculation, the structure of the exploding object, the location of its ignition point(s), its explosion mechanism, and its explosive dynamics would determine the products of nuclear burning and their distribution in the ejecta.
In the spirit of \cite{woosley2007}, we do not model these processes directly in our study; instead, we parametrize  the nucleosynthetic yields of burning. 
Spectral time series of \sneia\ indicate that the ejecta are compositionally stratified, with heavier elements located in the inner layers \citep[e.g.,][]{pereira13}. 
In models, such an abundance layering results from burning at higher densities producing heavier elements. 
The interior ejecta consists primarily of \Nifs\ and stable iron group elements (IGEs), the latter produced by electron capture or the neutron excess from metallicity. 
The exterior layers consist of intermediate mass elements (IMEs) and possibly unburned C/O.

Accordingly, our model possesses three layers: a \Nifs+IGE layer (95\% \Nifs, 5\% \Feff\ by mass), an IME layer, and an unburned C/O layer (50\% C, 50\% O by mass). 
For the IME layer, we base our abundances on the detailed nucleosynthesis calculations of \cite{shen17}, adopting the following mass fractions: $^{20}\mathrm{Ne}$: $4.486\times10^{-8}$, $^{24}\mathrm{Mg}$: $1.294\times10^{-5}$, $^{28}\mathrm{Si}$: $5.299\times10^{-1}$, $^{32}\mathrm{S}$: $3.303\times10^{-1}$, $^{36}\mathrm{Ar}$: $6.986 \times10^{-2}$, $^{40}\mathrm{Ca}$: $6.834 \times 10^{-2}$, $^{44}\mathrm{Ti}$: $5.268\times10^{-5}$, $^{48}\mathrm{Cr}$: $1.479\times10^{-3}$.
The masses of these layers are free parameters of the model, which we write as \fni, the fraction of the ejecta mass in the \Nifs+IGE layer, and \fco, the fraction of the ejecta mass in the C/O layer. 
The mass of the IME layer can be derived from the model parameters as $\mime=\mej(1-\fni-\fco)$.

There are a variety of mechanisms by which the abundances of elements in the ejecta can become smeared-out in velocity space (henceforth ``mixing"). 
Our model includes a mixing parameter $m$, a dimensionless real number that is related to \Nmix, the integer number of times to run a boxcar average of window size 0.02 \Msun\ over the ejecta in Lagrangian space, via $\Nmix = \mathrm{floor}(10^m)$.
In our scheme this parameter accomplishes the mixing of \Nifs\ and other species into the inner and outer regions of the atmosphere.

We generated a grid of \nmod\ \snia\ ejecta structures using the above scheme, each of which  is fully specified by the five parameters listed in Table \ref{tab:recap}. 
We drew the parameters of each model at random from independent uniform distributions over the ranges specified in Table \ref{tab:recap}, except for \mej, which was drawn from a broken uniform distribution. 
95\% of the models had $0.7 \leq \mej / \msun < 1.6$, and the remaining 5\% had $1.6 \leq \mej / \msun \leq 2.5$. 
All models in the grid were generated with the \q{simple}\ supernova atmosphere generator \citep{goldstein16}.

\section{Radiation Transport Simulations}
\label{sec:rad}

We used the time-dependent multi-wavelength Monte Carlo radiation transport code \MCcode\ \citep{sedona} to calculate light curves and spectra of the models.
Given a homologously expanding supernova ejecta structure, \MCcode\ calculates the full time series of emergent spectra.
Broadband light curves can be constructed by convolving the synthetic spectrum at each time with the appropriate filter transmission functions.
\MCcode\ includes a detailed treatment of gamma-ray transfer throughout the atmosphere to determine the instantaneous energy deposition rate from radioactive \Nifs\ and \Cofs\ decay \citep[for a detailed description of the gamma-ray transport, see][Appendix A]{sedona}. 
Other decay chains that can change the composition of trace isotopes, such as $\Crfe \rightarrow \Vfe \rightarrow \Tife$, are not treated in the present calculations.
Radiative heating and cooling rates are evaluated from Monte Carlo estimators, with the temperature structure of the ejecta assumed to be in radiative equilibrium.
See \cite{sedona} and \cite{roth15} for detailed code descriptions and verifications.

Several significant approximations are made in our \MCcode\ simulations, notably the assumption of local thermodynamic equilibrium (LTE) in computing the atomic level populations.
In addition, bound-bound line transitions are treated using the expansion opacity formalism (implying the Sobolev approximation; \citealt{jeffery95}).
In this formalism, the opacities of many spectral lines are represented in aggregate by a single effective opacity.
Although the \MCcode\ code is capable of a direct Monte Carlo treatment of NLTE line processes, due to computational constraints this functionality is not exploited in the large parameter survey here.

\begin{deluxetable}{cc}
\tablecaption{Model Parameters and Assumptions \label{tab:recap}}
\tablehead{\colhead{Parameter} & \colhead{Range}}
\startdata
\mej\ &  0.7 -- 2.5 \msun\\
$v_K$ & 8,000 -- 15,000 \kms \\
\fni\ & 0.1 -- 0.8\\
\fco\ & 0.00 -- 0.07\\
$m$ & 1.0 -- 2.5\\
\hline \multicolumn{2}{c}{Assumptions and Justifications}\\
\hline 
\multicolumn{2}{c}{Spherically symmetric ejecta (spectropolarimetry)}\\ 
\multicolumn{2}{c}{Object that explodes is a C/O WD (2011fe)} \\
\multicolumn{2}{c}{Decaying exponential density profile (explosion simulations)} \\ 
\multicolumn{2}{c}{Ejecta are stratified in three layers (spectroscopy)}\\
\multicolumn{2}{c}{Level populations assumed to be in LTE}\\
\multicolumn{2}{c}{Light curve powered exclusively by \Nifs\ decay chain}
\enddata
\end{deluxetable}

We treat the line source function in the two-level equivalent atom approximation, where a parameter $\epsilon$ sets the ratio of absorptive  opacity to total (i.e., absorptive plus scattering) opacity.
Previous LTE studies of SNe~Ia have found that using primarily absorptive lines ($\epsilon \approx 1$) reasonably captures the wavelength redistribution of photons. 
In complex atoms like the iron group species, redistribution via multiple fluorescence takes on an approximate thermal character \citep{pe2,pe1,blinnikov06}. 
For  strong IME line features, however, using $\epsilon = 1$ typically over-estimates the line emission component; for example, \cite{kasen06} showed that using a purely absorptive line source function for calcium over-predicts the emission in the Ca~II IR triplet features, which substantially influences the $I$-band light  curve. We therefore assume that lines for all ions with $Z \le 20$ are ``purely scattering" ($\epsilon_\mathrm{th} = 0$) whereas lines from ions near the iron group are ``purely absorptive'' (i.e.,  $\epsilon_{\rm th} = 1$). 

While the LTE approximation has been shown to produce a reasonable light curve predictions during the photospheric phases of SNe~Ia \citep[e.g.,][]{jack11} quantitative errors in the broadband magnitudes are expected on the order of 0.1 to 0.3~mag.
For this reason, our model peak magnitudes, colors and broadband (e.g., $B$-band) decline rates should be considered uncertain at this level.
The adoption of an alternative value $\epth \lesssim 1$ for the iron group lines shifts the location of the models in the WLR and color plots discussed below. In particular, choosing  $\epsilon \lesssim 1$ leads to slower B-band light curve decline rates, but does not significantly change the slope of the model relation or the level of dispersion. On the other hand, if \epth\ depends in a systematic way on temperature or density, this could lead to correlated errors that affect the slope of the model relation.

The two-level atom framework applied here is just one of several uncertainties that affect the radiative transfer calculations.
In addition, inaccuracy or incompleteness in the atomic line data can be a source of significant error. The inaccuracy of the LTE ionization assumption may also have significant consequences for the $B$-band light curves \citep{sedona}. At later times ($\ga 30$~days after $B$-band maximum) the NLTE effects become increasingly significant and the model calculations become unreliable.

The numerical gridding in the present calculations was as follows:
\emph{spatial:} 100 equally spaced radial zones with a maximum velocity of $4 \times 10^4$~\kms;
\emph{temporal:} 116 time points beginning at day~1 and extending to day 100 with logarithmic spacing $\Delta \log\,t = 0.1$ and a maximum time step of 1 day;
\emph{wavelength:} covering the range 150 -- 60,000~\AA\ with resolution $\Delta \log \lambda = 0.001$.
Atomic line list data were taken from the Kurucz CD~1 line list \citep{Kurucz_Lines}, which contains about 42 million lines.
A total of $\sim$$2.3\times10^7$ photon packets was used for each calculation, which allowed for acceptable signal-to-noise in the synthetic broadband light curves and spectra.

\section{Results}
\label{sec:correlations}

\begin{figure}
	\centering 
\includegraphics[width=0.5\textwidth]{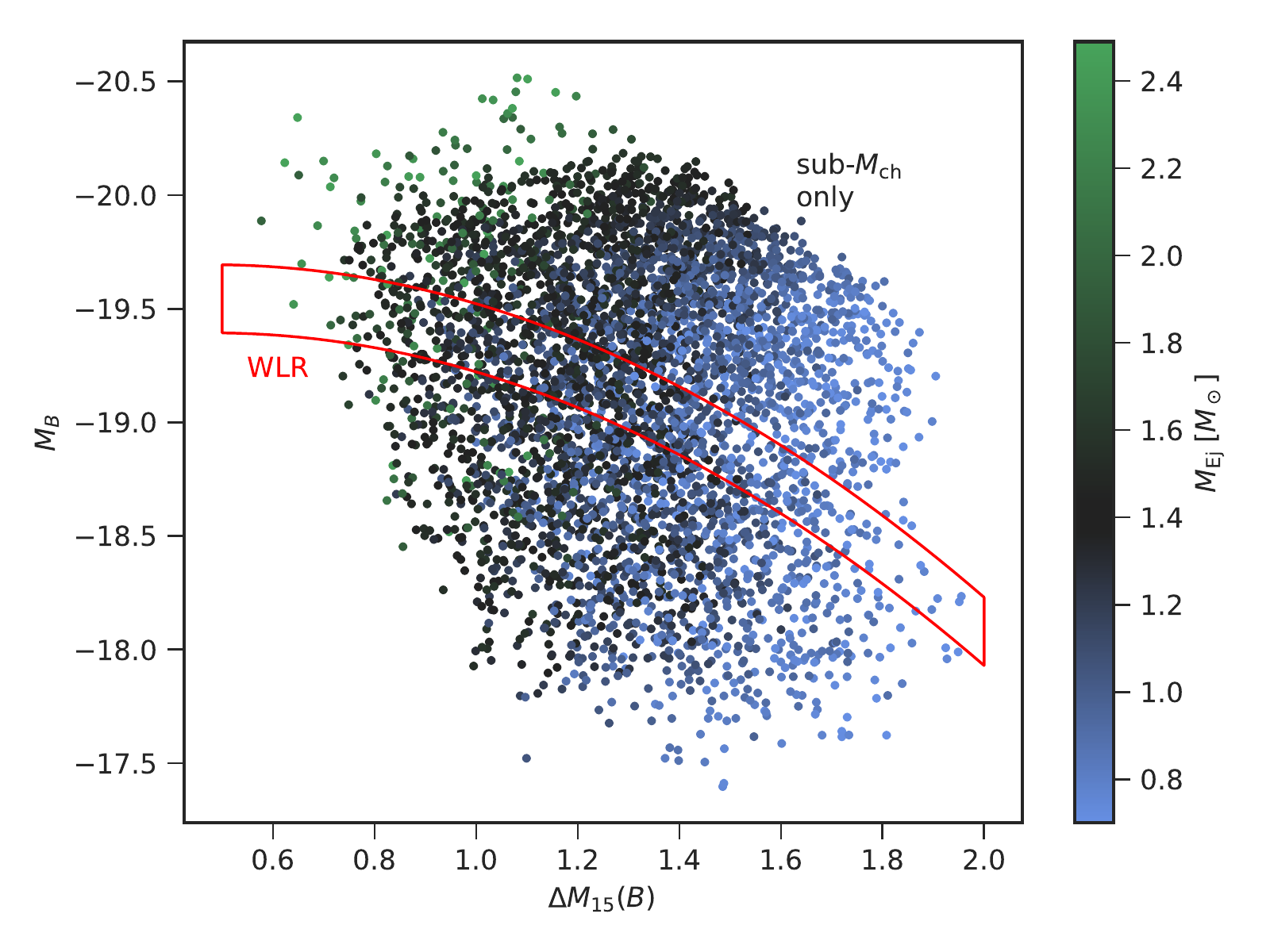}
    \caption{$M_B$ versus $\Delta M_{15}(B)$ for our entire grid of \nmod\ models.
    The relationship of \cite{phillips99} is overplotted as a red box.
    Models with $\dmfb \geq 1.55$ are definitively sub-\mch.
    Both \mch, sub-\mch, and even some super-\mch\ models are consistent with $0.8 < \dmfb < 1.55$. }
    \label{fig:phillips}
\end{figure}

Figure \ref{fig:phillips} shows the distribution of our models in \dmfb\ v.\ $M_B$ space, color coded by \mej, with magnitudes given in the Vega system.
The WLR of \cite{phillips99} is overplotted as a red box.
The models span the space of \sneia\ and beyond, including the sub-luminous 91bg and over-luminous 91T-like events, as well as many more unusual events that fall far from WLR and may have never been observed. 
As discussed in \cite{woosley2007}, this implies that Nature does not realize all conceivable ejecta structures, rather the existence of a WLR implies that the explosion physics of \sneia\ must correlate the ejecta parameters in a systematic way.

By eye, it is clear that the entire model set shows a strong trend between \dmfb, $M_B$, and \mej. 
Models with faster $B$-band decline rates have lower overall ejecta mass. 
Only models with $M < \mch$ can reproduce the fastest decline rates, $\dmfb > 1.55$.
The sub-\mch\ models can reproduce the full extent of the WLR $(0.8 < \dmfb < 2.0)$, whereas \mch\ models can only reproduce the bright end $(0.8 < \dmfb < 1.55)$.
Inclusion of NLTE effects \citep[e.g.,][]{blondin13,dessart14} is necessary to see if radiative transfer effects can change this conclusion, although  previous \snia\ light curve calculations suggest that NLTE effects tend to systematically decrease \dmfb\ compared to LTE calculations (compare, e.g., \citealt{blondin17} and \citealt{shen17}).

\begin{figure}
	\centering
    \includegraphics[width=1\columnwidth]{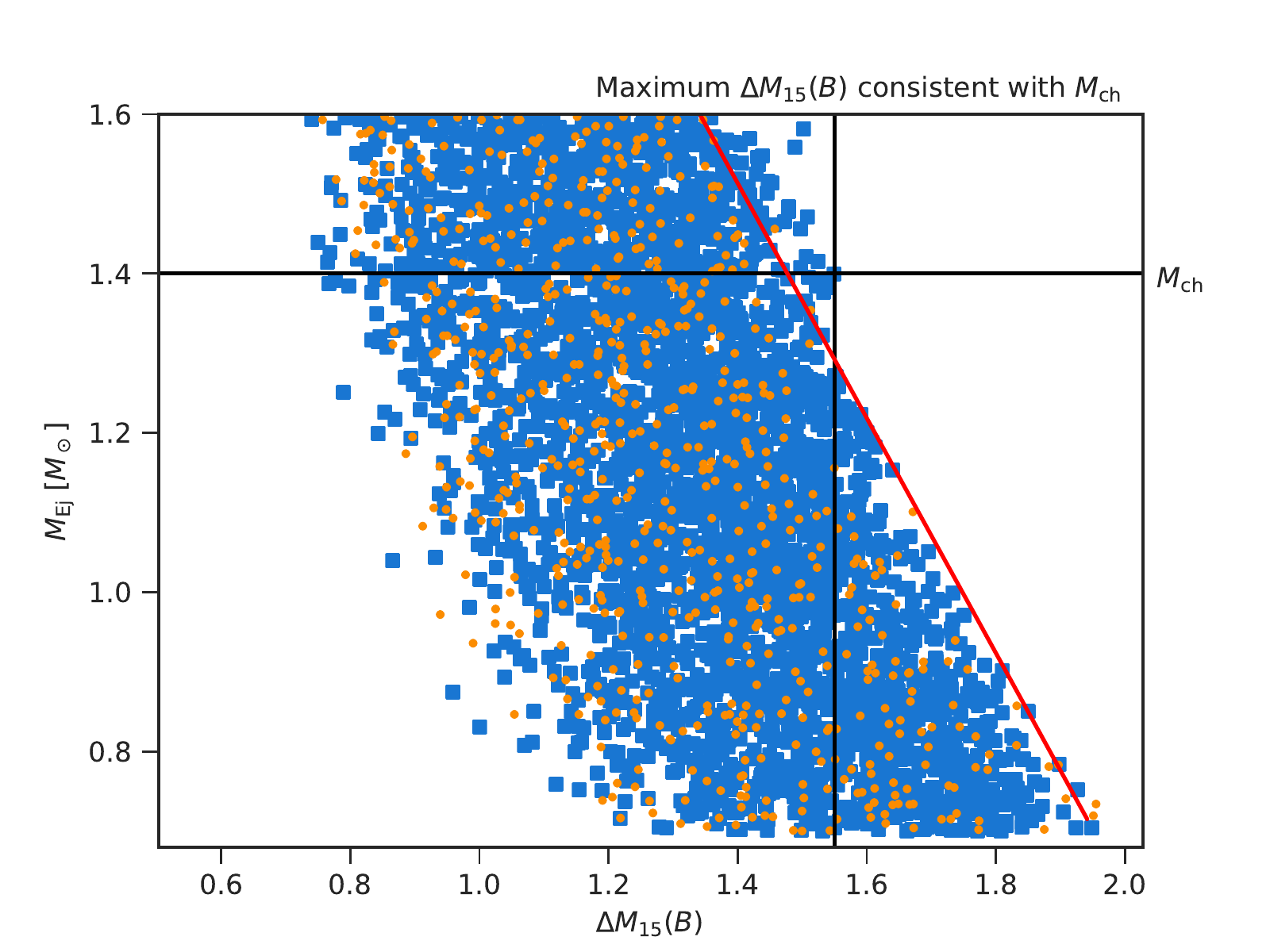}
    \caption{Correlation between \mej\ and \dmfb\ for models off the WLR (blue squares) and on the WLR (orange dots).
    In both cases, models with $\dmfb > 1.55$ cannot be explained by \mch\ progenitors.
    Many observed \sneia\ satisfy this criterion.
    A linear fit to the maximum ejected mass as a function of \dmfb\ is shown as a red line.}
    \label{fig:ej}
\end{figure}

Figure \ref{fig:ej} shows the correlation between \dmfb\ and \mej\ for models on and off the WLR (for clarity, super-\mch\ models have been excluded from the Figure). 
At $\dmfb=1.55$, models on the grid cease to be consistent with \mch.
Many known \sneia\ have $\dmfb > 1.55$; for an in-depth study of supernovae of this class see \cite{taubenberger08}.
We conclude that sub-\mch\ progenitors constitute at least some and potentially the bulk of observed of \sneia.

Evidence for sub-\mch\ \sneia\ has previously been derived from empirical analyses of pseudobolometric SN~Ia light curves \citep{stritzinger06,scalzo14a,scalzo14b,dhawan17,wygoda17}.
\cite{heringer17},
using LTE spectrum synthesis calculations, found that a common explosion mechanism can account for both sub-luminous and normal \sneia. 
Our work suggests that this common mechanism is likely sub-\mch. 
\cite{blondin17}, using NLTE light curve calculations for a set of 12 detonation models similarly found that only sub-\mch\ models could reproduce the fast-declining light curves. Our extensive model survey indicates that this conclusion is robust and holds over a much more general model parameter space that is not tied to a particular explosion scenario.

Interestingly, we note that super-luminous \sneia\ such as  SN 2003fg, SN 2006gz, SN 2007if, and SN 2009dc,  all fall squarely in the super-\mch\ zone of Figure \ref{fig:phillips} \citep{scalzo10}, suggesting that such events do indeed require super-\Mch\ ejecta, unless they are not spherically symmetric, or an additional source of luminosity (e.g., interaction) contributes to the brightness \citep{maeda10,silverman13}.

\subsection{Light Curves and Empirical Fitting Functions}

\begin{figure*}[p]
\begin{minipage}{1\columnwidth}
	\centering
    \includegraphics[width=1\columnwidth]{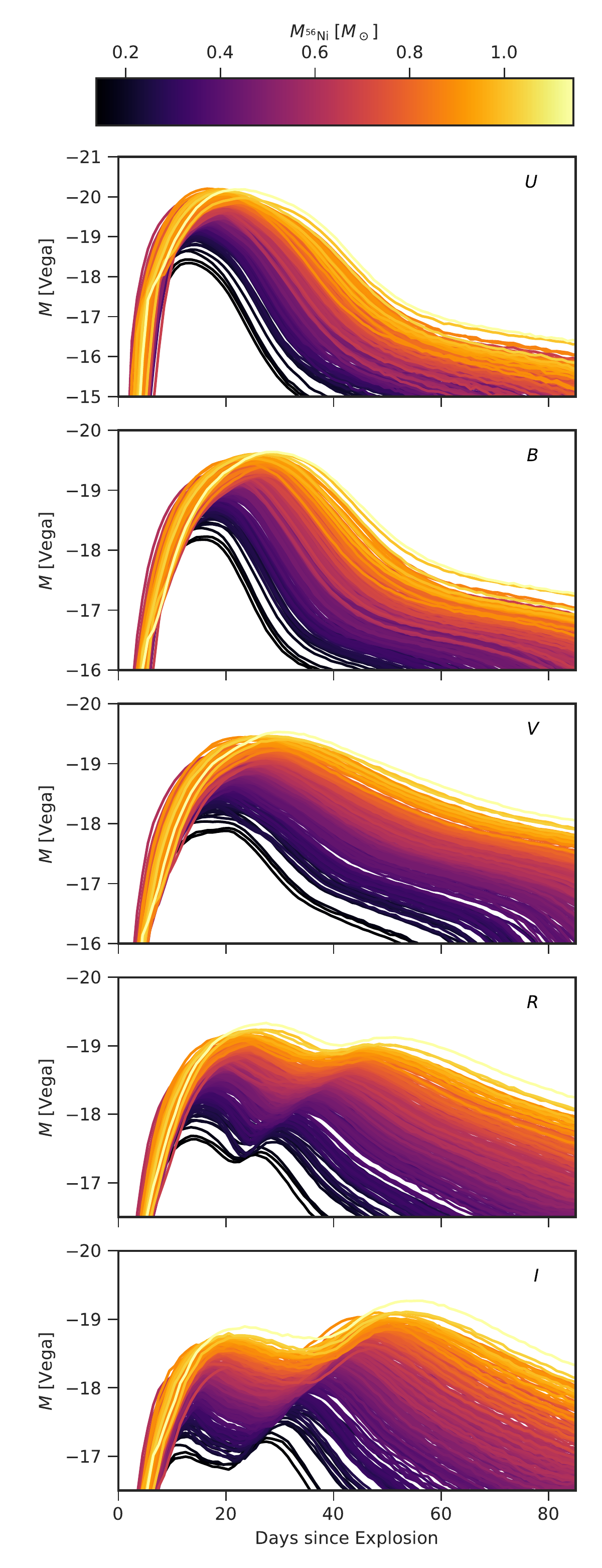}
    \caption{\MCcode\ light curves for the models that lie along the WLR (red boxed region in Figure \ref{fig:phillips}), color-coded by \mni.
    There is a strong correlation between light curve shape and \mni\ in all bands.}
    \label{fig:lcs}
\end{minipage}
\hspace{2mm}
\begin{minipage}{1\columnwidth}
	\centering
\includegraphics[width=1\columnwidth]{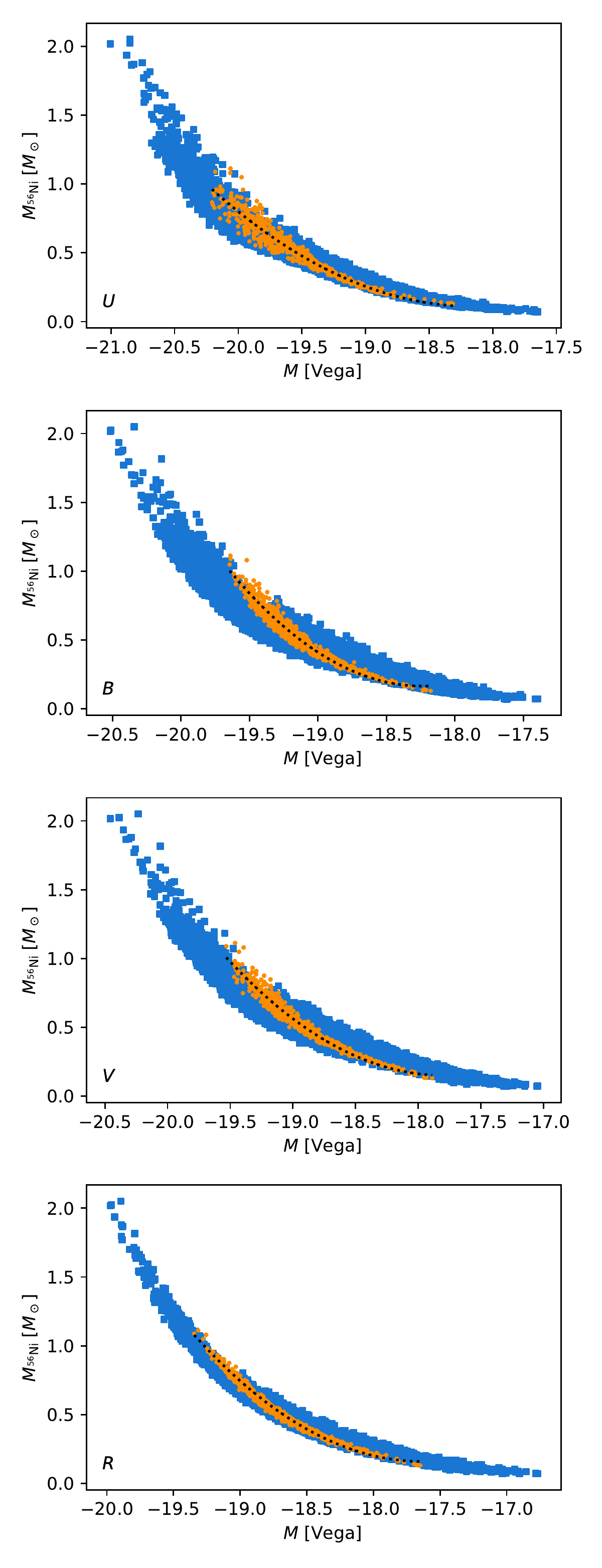}
    \caption{Correlation between \mni\ and peak brightness for models off the WLR (blue squares) and on the WLR (orange dots).
    The quadratic fitting functions in Equations \ref{eq:u} -- \ref{eq:r} are plotted as black dashed lines in each panel.}
    \label{fig:ni}
\end{minipage}
\end{figure*}

To make the results of our large simulation grid useful to observers, we provide simple fitting functions that map empirical properties of \sneia\ to physical parameters of our models.  
Figure \ref{fig:lcs} shows the $UBVRI$ light curves of the models that lie along the WLR of Figure \ref{fig:phillips}, color-coded by \mni. 
These light curves exhibit the characteristic broadband evolution of \sneia, validating the model parametrization described in Section \ref{sec:explosion}.
The over-luminous secondary maximum in $R$ and $I$ bands is a well-known issue related to the assumptions in the transfer (see Section \ref{sec:rad}). 
The light curves show a strong correlation between \mni\ and peak brightness in all bands; this is illustrated in Figure \ref{fig:ni}. 
The relations are quadratic and given in the Appendix.

Figure \ref{fig:ej} shows that there is a linear relationship between \dmfb\ and the maximum ejected mass of an \snia. 
We fit a linear relationship to the data in the Figure, where it is displayed as a red line. 
We report the equation describing the relationship in the Appendix.
We did not find a significant correlation between the mixing parameter $m$ and the observable properties of the models.  
Goldstein \& Kasen (in preparation) will present a more thorough analysis of the entire model set, and release it to the public. 

\section{Discussion}
\label{sec:discussion}

Although our model space is broad, it is not exhaustive---we have imposed physical constraints that limit certain models from the grid.
One potentially important example is that, unlike some previous 1D investigations, we have not considered models with large cores of stable iron group elements.
Such cores can help increase the decline rates of models by moving the \Nifs\ closer to the surface,  reducing the diffusion time and providing  lines that blanket the $B$-band over time. 

We do not include a large central core of stable IGEs in our models for two reasons.
First, such  cores appear to be the result of artificial burning fronts prescribed in  parameterized 1D delayed-detonation models. 
In more realistic three-dimensional simulations, the deflagration phase produces buoyant plumes that smear out the compositional structures of the central regions, leaving no stable core \citep[e.g.,][]{seitenzahl13}. 
Other models of SNe Ia (e.g., sub-\Mch\ double detonations, violent WD mergers, WD collisions) likewise do not produce a stable iron core.

Second, NLTE calculations show that a large core of stable iron produces nebular line profiles that disagree with observations---notably a flat top to the [Co III] feature at 5888\AA\ \citep{botyanszki17}.
Observations of flat-topped profiles of Co lines in near-IR spectra of SN 2003du had been the main observational evidence given by \cite{hoeflich04}  for the necessity of a stable iron core in models. 
However, the signal to noise ratio of the nebular spectrum of SN 2003du was low, and more recent samples of high signal-to-noise \snia\ nebular spectra have definitively exhibited rounded profiles (e.g.,  Maguire et al. in preparation). 

Another potentially important effect is that we only consider exponential density profiles in our grid.
Ejecta models with flat or slowly-declining central density profiles have smaller central densities for a given kinetic energy and mass than the exponential profiles used here, leading to shorter diffusion times and faster-evolving light curves. 
To determine if using a density profile with a flat inner region  could enable  our \mch\ models to account for fast-declining \sneia,  we selected 450 models (10\%) at random from our grid  and re-ran them using a broken power law density profile
\begin{equation} 
\rho(v) \propto  \begin{cases} 
      (v/v_t)^{-\delta} & v < v_t \\
      (v/v_t)^{-n} & v \geq v_t
   \end{cases},
\end{equation}
with a flat inner region $(\delta = 0)$ and a steeply declining outer region $(n = 10)$ instead of an exponential, with $v_t$ set by $E_K$ and \Mej\ following \cite{kasen10}.
We kept all other model parameters (mass, composition, mixing, and kinetic energy) the same. 

We find that although flattening the inner density profile increases the decline rates of \Mch\ models, the models still cannot reproduce the fastest-declining \sneia. 
The exponential and broken power law models differ quantitatively by $\sim$0.2 mag in the \dmfb\ value at which they cease to be consistent with \Mch\ ($\dmfb = 1.75$ for the broken power law models versus $\dmfb = 1.55$ for the exponential models).
But the power law \mch\ models that have relatively fast decline rates $(\dmfb\ > 1.6)$ require unrealistically high kinetic energies $(2 \times 10^{51} \;\mathrm{erg} < E_K  < 3 \times 10^{51} \;\mathrm{erg}$) and \Nifs\ yields $(\mni > 1 \msun)$. 
Thus, although some \mch\ power law models can get close to fast-declining \sneia, they can only do so with extreme parameters that cause other problems, suggesting that such models are unlikely.

Our main results are in conflict with those of  \cite{hoeflich17}, who make the case that with the appropriate choice of parameters they can account for the photometric behavior of both typical \sneia\ and 91bg-like events using \mch\ delayed-detonation models. 
An important factor that likely allows \cite{hoeflich17} to achieve fast declining \Mch\ models is that their low \Nifs\ models have a large amount (0.2 -- 0.3 \Msun) of stable iron at the very center. 
We note that the results of \cite{hoeflich17} are in significant tension with the results of \cite{blondin17}. 
Running similar low-\Nifs\ \Mch\ ejecta structures (DDC25 and 08) through NLTE radiation transport, \cite{hoeflich17} obtain \dmfb\ values that are larger than those of \cite{blondin17} by almost 1 mag. 
While resolving this discrepancy is outside the scope of the present paper, we note that the 3D calculations of \cite{sim13} also suggest that \mch\ models with low \Nifs\ yields do not have fast decline rates.

\section{Conclusion}
\label{sec:conclusion}
In this work we have used an extensive suite of \nmod\ detailed radiation transport simulations  to show that fast-declining \sneia\ come from sub-\mch\ progenitor systems. 
We find that sub-\mch\ and \mch\ \sneia\ can reproduce the bright end of the WLR $(0.8 < \dmfb < 1.55)$, whereas only sub-\mch\ \sneia\ can reproduce the faint end $(\dmfb > 1.55)$. 
In the era of big data, systematic parameter studies such as the one presented here will be  useful  for understanding the  physics of transients. 
In future papers, we will use this technique to further illuminate the nature of \sneia, and we will expand the grid to further test the consistency of \mch\ models with \snia\ light curves and spectra. 

\appendix
\section{Functions Relating Physical Parameters of \sneia\ to Observable Quantities}
The following functions can be used to estimate the \Nifs\ mass of an \snia\ given its absolute magnitude in the rest-frame $U,B,V$ or $R$ bands (Vega system). 
The first set of functions gives a lower bound on the \Nifs\ mass, the second set gives a median, and the third set gives an upper bound. 
The functions are second-degree polynomial fits to the 2.5, 50, and 97.5th percentiles of magnitude-binned WLR (orange) points in Figure \ref{fig:ni}.
\begin{align}
[\mni / \msun]_\mathrm{min} &= 0.16263 M_U^2 + 5.89704 M_U + 53.57331\\
&= 0.43270 M_B^2 + 15.85282 M_B + 145.37146\\
&= 0.27958 M_V^2 + 9.98986 M_V + 89.39890\\
&= 0.33515 M_R^2 + 11.87870 M_R + 105.42028
\end{align}

\begin{align}
[\mni / \msun]_\mathrm{med} &= 0.20054 M_U^2 + 7.27769 M_U + 66.13664\label{eq:u}\\
&= 0.42325 M_B^2 + 15.43821 M_B + 140.94416\label{eq:b}\\
&= 0.28458 M_V^2 + 10.12929 M_V + 90.28593\label{eq:v}\\
&= 0.31409 M_R^2 + 11.07772 M_R + 97.83555\label{eq:r}
\end{align}

\begin{align}
[\mni / \msun]_\mathrm{max} &= 0.25592 M_U^2 + 9.32807 M_U + 85.11384\\
&= 0.40649 M_B^2 + 14.71865 M_B + 133.37570\\
&= 0.30942 M_V^2 + 10.98176 M_V + 97.57820\\
&= 0.29679 M_R^2 + 10.41115 M_R + 91.45071
\end{align}

The following function gives the maximum ejected mass of an \snia\ in terms of \dmfb. 
It is a fit to the upper boundary of the blue points in Figure \ref{fig:ej}, where it is shown as a red line.
\begin{equation}
M_{\mathrm{Ej, max}} = -1.47 \dmfb + 3.57
\end{equation}

\acknowledgements
DAG gratefully recognizes Rollin Thomas for years of guidance and mentorship---they were essential to this work.
Ken Shen deserves thanks for sharing the results of his nucleosynthesis calculations and Peter Nugent for useful discussions.
The authors acknowledge an anonymous referee for comments that improved the paper. 
This work was supported in part by the U.S. Department of Energy, Office of Science, Office of Nuclear Physics, under contract number DE-AC02-05CH11231 and DE-SC0017616, and by a SciDAC award DE-SC0018297. 
DK acknowledges support from the Gordon and Betty Moore Foundation through Grant GBMF5076.  
This research used resources of the National Energy Research Scientific Computing Center, a DOE Office of Science User Facility supported by the Office of Science of the U.S. Department of Energy under Contract No. DE-AC02-05CH11231.
This research made use of NASA's Astrophysics Data System.

\ \newline

\bibliography{emulator.bib}

\end{document}